\documentclass[aps,prl,twocolumn,amsmath,amssymb,nofootinbib,superscriptaddress,floatfix,reprint,shortbibliography]{revtex4-2}
\usepackage{latexsym}
\usepackage{amsmath}
\usepackage{amsfonts}
\usepackage{amssymb}
\usepackage{bm}
\usepackage{color}
\usepackage{txfonts}
\usepackage{float}
\usepackage{url}
\usepackage[colorlinks=true, urlcolor=blue, linkcolor=blue, citecolor=blue, pdftex]{hyperref}
\usepackage{ulem}
\usepackage{graphicx}
\usepackage{extpfeil}
\usepackage{subfigure}
\usepackage{physics}
\usepackage{setspace}
\DeclareUnicodeCharacter{2212}{-}

\begin{document}

\newcommand{\fig}[2]{\includegraphics[width=#1]{#2}}
\newcommand{\la}{{\langle}}
\newcommand{\ra}{{\rangle}}
\newcommand{\dg}{{\dagger}}
\newcommand{\upa}{{\uparrow}}
\newcommand{\dna}{{\downarrow}}
\newcommand{\ab}{{\alpha\beta}}
\newcommand{\ias}{{i\alpha\sigma}}
\newcommand{\ibs}{{i\beta\sigma}}
\newcommand{\hH}{\hat{H}}
\newcommand{\hn}{\hat{n}}
\newcommand{\hc}{{\hat{\chi}}}
\newcommand{\hU}{{\hat{U}}}
\newcommand{\hV}{{\hat{V}}}
\newcommand{\cH}{{\mathcal{H}}}
\newcommand{\br}{{\bm{r}}}
\newcommand{\bk}{{{\bm{k}}}}
\newcommand{\bq}{{{\bm{q}}}}
\newcommand{\bp}{{{\bm{p}}}}
\def\gsim{~\rlap{$>$}{\lower 1.0ex\hbox{$\sim$}}}
\setlength{\unitlength}{1mm}
\newcommand{{\vhf}}{$\chi^\text{v}_f$}
\newcommand{{\vhd}}{$\chi^\text{v}_d$}
\newcommand{{\vpd}}{$\Delta^\text{v}_d$}
\newcommand{{\ved}}{$\epsilon^\text{v}_d$}
\newcommand{{\vved}}{$\varepsilon^\text{v}_d$}
\newcommand{\pprl}{Phys. Rev. Lett. \ }
\newcommand{\pprb}{Phys. Rev. {B}}
\newcommand{\LNO}{La$_6$Ni$_5$O$_{12}$ }

 \title{$d$-wave superconductivity in quintuple-layer bulk superconductor \LNO}
 \author{Zhan Wang}
\thanks{These authors contributed equally to this work.}
 \affiliation{Beijing National Laboratory for Condensed Matter Physics and,\\ Institute of Physics, Chinese Academy of Sciences, Beijing 100190, China}

 \author{Yuxin Wang}
\thanks{These authors contributed equally to this work.}
\affiliation{Kavli Institute for Theoretical Sciences, University of Chinese Academy of Sciences,
	Beijing, 100190, China}
    
\author{Hui-Ke Jin}
\email{jinhk@shanghaitech.edu.cn}
\affiliation{School of Physical Science and Technology, ShanghaiTech University, Shanghai 201210, China}

 \author{Kun Jiang}
 \email{jiangkun@iphy.ac.cn}
 \affiliation{Beijing National Laboratory for Condensed Matter Physics and,\\ Institute of Physics, Chinese Academy of Sciences, Beijing 100190, China}
 \affiliation{School of Physical Sciences, University of Chinese Academy of Sciences, Beijing 100190, China}

\author{Jiangping Hu}
\email{jphu@iphy.ac.cn}
\affiliation{New Cornerstone Science Laboratory, Beijing National Laboratory for Condensed Matter Physics and Institute of Physics, Chinese Academy of Sciences, Beijing 100190, China}

\date{\today}

\begin{abstract}
    The recent realization of bulk superconductivity in the quintuple-layer nickelate La$_6$Ni$_5$O$_{12}$ opens new opportunities to probe the pairing symmetry of the reduced Ruddlesden--Popper (RP) nickelate family. Using an effective five-layer Hubbard model and variational Monte Carlo calculations, we find that a spin-singlet $d_{x^2-y^2}$-wave state is energetically favored, while extended $s$-wave pairing is strongly suppressed. Based on the optimized superconducting state, we further calculate several experimentally accessible observables. The tunneling density of states exhibits a characteristic V-shaped low-energy spectrum, the superfluid density shows a linear low-temperature depletion, the Knight shift decreases below $T_c$, and the spin-lattice relaxation rate follows $1/T_1T\propto T^2$, together providing consistent signatures of a clean nodal superconductor. Extending the same framework to related reduced RP nickelates, we find a robust $d$-wave tendency in infinite-layer LaNiO$_{2}$, while trilayer La$_4$Ni$_3$O$_{8}$ at its nominal filling lies close to the boundary between $d$-wave superconductivity and a non-superconducting state. These results establish bulk La$_6$Ni$_5$O$_{12}$ as a promising platform for determining the pairing symmetry of nickelate superconductors and suggest a common $d_{x^2-y^2}$-wave pairing tendency across the reduced RP nickelate family.
\end{abstract}

\maketitle

Superconductivity in reduced Ruddlesden--Popper (RP) nickelates has opened a new route to unconventional pairing beyond the cuprates~\cite{Li2019,Li2020,Zeng2020,Pan2022,Pan2026,Chow2025,2026Lib,NomuraReview2022,WangReview2024,2025Chene,Hepting_review,Parzyck2025,Sahib2025,Dong2025,Si2020Hydrogen,Hepting2020}. A central question is the symmetry of the superconducting order parameter~\cite{NomuraReview2022,WangReview2024,2025Chene}. Theoretical studies of infinite-layer \(R_{1-x}A_x\mathrm{NiO}_2\) (112) have predominantly favored spin-singlet \(d_{x^2-y^2}\)-wave pairing~\cite{Anisimov1999,LeePickett2004,Jiang2019,Nomura2019,Hu2019,Ryee2020,Jiang2020Spin,Zhang2020S,Gu2020,Werner2020,Lechermann2020,Botana2020,Wu2020, Sakakibara2020, Kitatani2020, Wang2020, Adhikary2020, Kreisel2022, Chen2025}, and recent experiments have provided evidence for a nodal superconducting gap~\cite{Cheng2024, 2025Hwangi,2023Ariando,Ranna2025}. Yet a definitive characterization of the pairing state remains difficult because superconductivity in these systems has so far been realized primarily in epitaxial thin films~\cite{Li2019,Li2020,Zeng2020,Pan2022}, strongly restricting conventional bulk probes such as nuclear magnetic resonance (NMR), penetration depth, and thermodynamic measurements~\cite{NomuraReview2022,WangReview2024,2025Chene}.

\begin{figure}
    \centering
    \includegraphics[width=\linewidth]{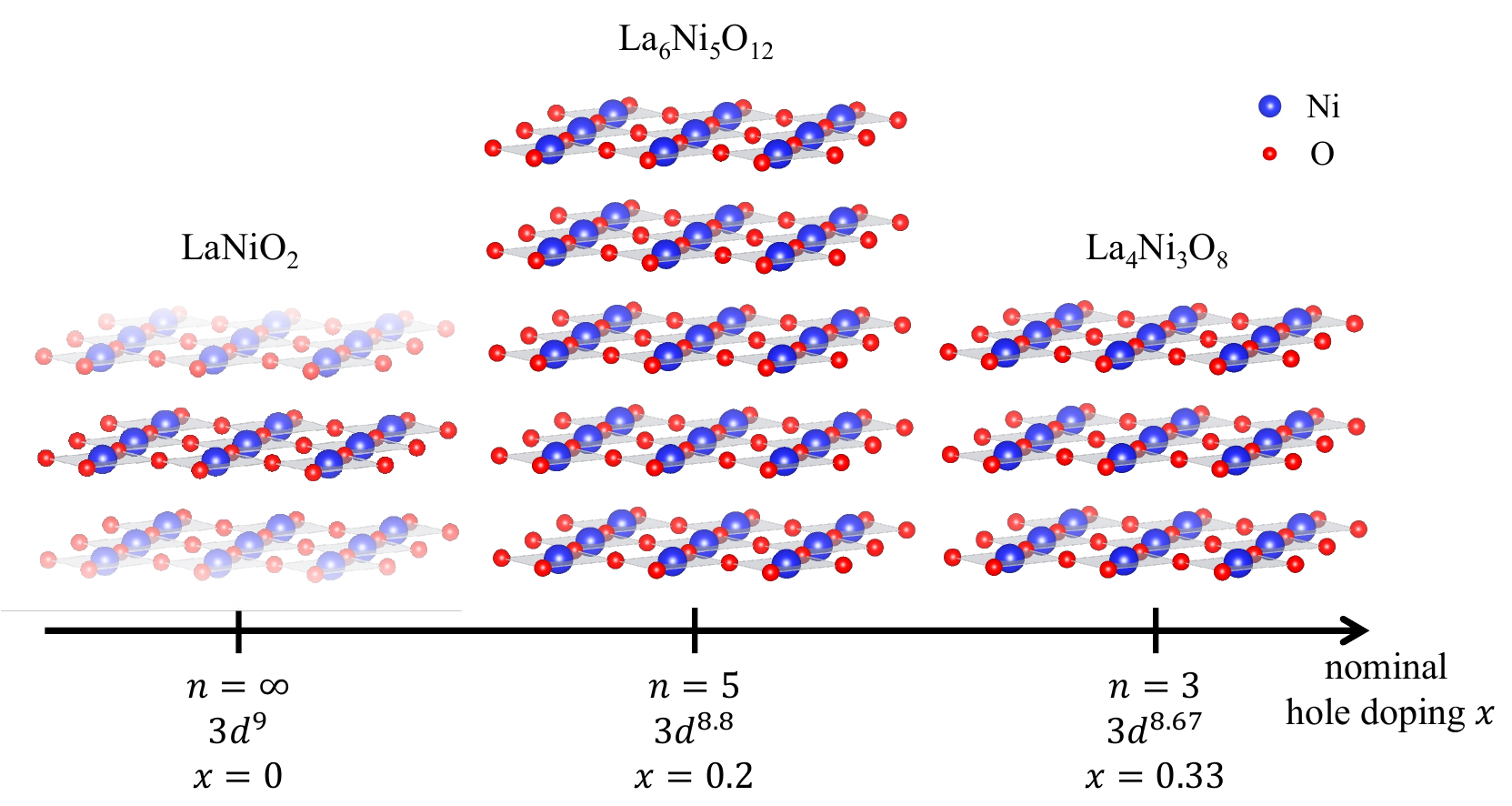}
    \caption{Schematic NiO$_2$-layer structures of the reduced RP nickelate family including the La\(\mathrm{Ni}\mathrm{O}_{2}\), La\(_6\mathrm{Ni}_5\mathrm{O}_{12}\), and La\(_4\mathrm{Ni}_3\mathrm{O}_{8}\). From left to right, the number of consecutive NiO$_2$ layers decreases from $n=\infty$ to $n=5$ and $n=3$. The corresponding nominal Ni configurations are $d^9$, $d^{8.8}$, and $d^{8.67}$, giving hole concentrations $x=0$, $0.2$, and $1/3$ relative to $d^9$, respectively. Superconductivity has been established in hole-doped infinite-layer thin films~\cite{Li2019}, with onset temperatures approaching $40$ K in Sm-based compounds~\cite{Chow2025, 2026Lib}, and in bulk 6512~\cite{2026Chen}, whereas signatures have also been reported in 438 single crystals~\cite{Ahn2026}.}
    \label{fig:model}
\end{figure}

The recent realization of bulk superconductivity in the quintuple-layer nickelate \(R_6\mathrm{Ni}_5\mathrm{O}_{12}\) (6512) provides a new opportunity to overcome this limitation~\cite{2026Chen}. Finite-layer nickelates provide a natural extension of the infinite-layer compounds. As illustrated in Fig.~\ref{fig:model}, the nominal Ni electron count in the \(R_{n+1}\mathrm{Ni}_n\mathrm{O}_{2n+2}\) series is \(d^{9-1/n}\), connecting the \(d^9\) infinite-layer limit to the formal \(d^8\) endpoint at \(n=1\), represented by \(\mathrm{La}_2\mathrm{NiO}_4\)~\cite{Pan2022,Worm2022,LaBollita2022}. Relative to the \(d^9\) configuration, this corresponds to a nominal hole doping of \(x=1/n\). Undoped 112 compounds have a nominal hole doping $x=0$, with doped holes introduced by self-doping from rare earth $5d$ orbitals~\cite{Jiang2019,Nomura2019} and external doping from rare earth substitutions. As the number of layers decreases, the nominal hole doping increases, reaching \(x=0.2\) in 6512~\cite{Pan2022,Pan2026} and \(x=1/3\) in trilayer \(R_4\mathrm{Ni}_3\mathrm{O}_8\) (438). Recent reports of ambient-pressure superconducting signatures in substituted 438 single crystals~\cite{Ahn2026} further motivate a comparison across the reduced RP nickelate family.

The nominal filling of 6512 is particularly appealing because it lies close to the superconducting doping regime of infinite-layer nickelates~\cite{Li2020,Zeng2020,Pan2022,Pan2026}. Several electronic-structure calculations support a low-energy description of quintuple-layer nickelates dominated by the Ni \(d_{x^2-y^2}\) orbital and find close similarities to hole-doped infinite-layer nickelates at comparable nominal filling~\cite{Worm2022,LaBollita2022,Botana2022,Hausoel2025}. An alternative multiorbital picture involving substantial Ni \(d_{z^2}\) flat-band contributions near the Fermi level has also been proposed~\cite{Lechermann2022}. Bulk 6512 therefore provides an experimentally accessible setting in which the pairing physics of reduced RP nickelates can be examined using complementary bulk-sensitive probes.

Previous theoretical studies have described finite-layer nickelates using effective single-band Hubbard models and obtained \(d\)-wave superconductivity mediated by antiferromagnetic spin fluctuations~\cite{Worm2022,Hausoel2025}. Complementary calculations based on a five-layer $t-J$ model have also found dominant $d_{x^2-y^2}$-wave pairing and determined layer-resolved superconducting gaps within a large-$N$ saddle-point treatment~\cite{Pan2022}. These results motivate an examination of pairing in the five-layer Hubbard model, together with predictions for bulk-sensitive experimental probes. 

Here, we address these questions using variational Monte Carlo (VMC) calculations for a realistic five-layer model of 6512. We find that a spin-singlet \(d_{x^2-y^2}\)-wave state is energetically favored and examine its layer-resolved pairing amplitudes. Based on this correlated superconducting state, we further calculate the quasiparticle tunneling density of states (TDOS), superfluid density, Knight shift, and nuclear spin-lattice relaxation rate $1/T_1T$, providing direct predictions for bulk experiments. In addition, we compare 6512 with trilayer 438 and infinite-layer 112, examining how pairing tendencies differ across nickelate compounds with different layer numbers. Our results identify bulk 6512 as a particularly valuable platform for probing the pairing symmetry of nickelate superconductors and for connecting superconductivity across the broader reduced RP nickelate family.

\begin{figure}
    \centering
    \includegraphics[width=\linewidth]{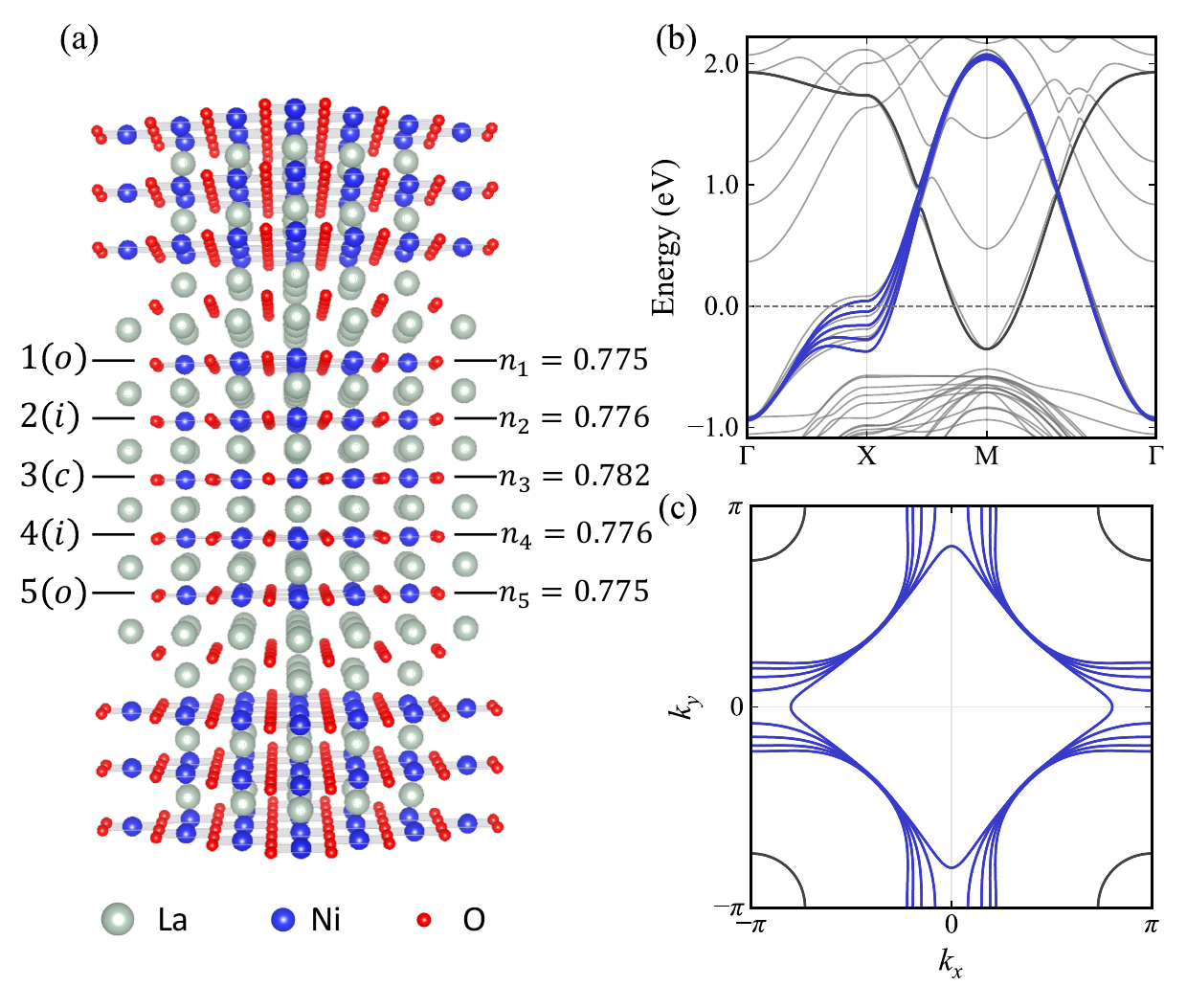}
    \caption{(a) Crystal structure of \LNO. The layer index $\ell$ and the layer types $(o,i,c)$, with mirror-related layers sharing the same type, are labeled explicitly. The layer-resolved electron fillings $n_\ell$ obtained from the TB model are indicated on the right. (b) Band dispersion of the six-band TB model. The five predominantly Ni $3d_{x^2-y^2}$ bands are shown in blue, and the rare-earth-derived band is shown in black. DFT reference  bands are shown in light gray. (c) Fermi surface of the six-band TB model, using the same color scheme as in (b).} 
    \label{fig:normal}
\end{figure}

The crystal structure of the quintuple-layer 6512 compound consists of five consecutive NiO\(_2\) planes formed by topotactic reduction, as illustrated in Fig.~\ref{fig:normal}(a). We label the layers as outer (\(o\))--inner (\(i\))--center (\(c\))--inner (\(i\))--outer (\(o\)), with mirror-related layers being equivalent. Density functional theory (DFT) calculations show that five bands predominantly derived from the Ni \(d_{x^2-y^2}\) orbitals, together with an additional La-derived band, cross the Fermi level, as depicted in Fig.~\ref{fig:normal}(b). We therefore construct a two-dimensional five-band tight-binding (TB) model for the Ni \(d_{x^2-y^2}\) sector through Wannierization, while treating the La-derived band as a self-doping reservoir that modifies the filling of the Ni \(d_{x^2-y^2}\) bands. Details of DFT calculations are provided in Supplemental Material~\ref{appendix:dft}.

More precisely, the TB Hamiltonian $\cH_t$ is written in momentum space as
\begin{align}
    \cH_t =& \sum_{\bk,\sigma}\left[\sum_{\ell=1}^{5}\varepsilon_{\ell,\bk}c_{\bk\ell\sigma}^\dagger c_{\bk\ell\sigma}+\sum_{\langle\ell\ell^\prime\rangle} \left(\tau_{\ell\ell',\bk}c_{\bk\ell\sigma}^\dagger c_{\bk\ell'\sigma}+{\rm h.c.}\right)\right].
    \label{eq:Ht}
\end{align}
Here $c_{\bk\ell\sigma}^\dagger$ creates an electron on layer $\ell$ with momentum $\bk$ and spin $\sigma$, and $\langle\ell\ell'\rangle$ denotes the summation over adjacent layers. In the following, we label each layer from $1$ to $5$ as indicated in Fig.~\ref{fig:normal}(a). Consequently, the first term in Eq.~\eqref{eq:Ht} describes the intralayer dispersion, while the second describes hopping between adjacent layers.

For each layer \(\ell\), the intralayer dispersion is written as
\begin{equation}
    \varepsilon_{\ell,\bk}
    =
    \epsilon_\ell
    +
    \sum_{m=1}^{3} t_{m\ell}\gamma_{m\bk},
\end{equation}
where \(\epsilon_\ell\) is the layer-dependent onsite energy, \(t_{m\ell}\) denotes the \(m\)th nearest neighbor intralayer hopping. The corresponding lattice harmonics are $\gamma_{1\bk}=2(\cos k_x+\cos k_y)$, $\gamma_{2\bk}=4\cos k_x\cos k_y$, and $\gamma_{3\bk}=2(\cos 2k_x+\cos 2k_y)$.
Along the out-of-plane direction, we retain only hopping between adjacent NiO\(_2\) layers, parameterized as
\begin{equation}
    \tau_{\ell\ell',\bk} = -2t_{\perp}^{\ell\ell'} (\cos k_x-\cos k_y)^2,
\end{equation}
which has the same momentum dependence commonly used for multilayer cuprates \cite{Anderson1995,FengDL_PhysRevLett.86.5550}. The complete set of tight-binding parameters is listed in Table~\ref{tab:TB}. The noninteracting band structure and Fermi surfaces are shown in Figs.~\ref{fig:normal}(b) and \ref{fig:normal}(c), respectively. The five Ni-derived Fermi-surface sheets resemble those of a cuprate-like band at different hole concentrations. The layer-resolved fillings are indicated in Fig.~\ref{fig:normal}(a). The five Ni-derived bands have a total electron filling of approximately $3.9$ per in-plane unit cell, corresponding to $22\%$ hole doping.

\begin{table}[!t]
    \centering
    \begin{tabular}{c|ccccc}
        \hline
        & $\epsilon_\ell$ & $t_{1\ell}$ & $t_{2\ell}$ & $t_{3\ell}$ & $t_\perp^{\ell\ell'}$ \\
        \hline
        outer
        & $0.438$ & $-0.374$ & $0.091$
        & $-0.047$ & $t_\perp^{oi}=0.014$ \\
        inner
        & $0.414$ & $-0.369$ & $0.091$
        & $-0.045$ & $t_\perp^{ic}=0.015$ \\
        center
        & $0.413$ & $-0.370$ & $0.090$
        & $-0.045$ & --- \\
        \hline
    \end{tabular}
    \caption{Tight-binding parameters of the five-layer model derived from DFT calculations, in units of eV. The last column lists outer--inner and inner--center hopping, with $t_\perp^{oi}=t_\perp^{io}$ and $t_\perp^{ic}=t_\perp^{ci}$.}
    \label{tab:TB}
\end{table}

For the interaction term, we include an onsite Hubbard repulsion \(U\) on each NiO\(_2\) layer. 
The full Hamiltonian is therefore
\begin{equation}
    \cH=\cH_t+U\sum_{i\ell} n_{i\ell\uparrow} n_{i\ell\downarrow},
    \label{eq:Hubbard}
\end{equation}
where $n_{i\ell\sigma}$ denotes the electron number operator at site $i$ in layer $\ell$ with spin $\sigma$. The total carrier concentration is controlled by fixing the number of doped holes \(N_h\).

We investigate the ground state of the multilayer Hubbard model using a Gutzwiller--Jastrow wave function,
\begin{equation}
    |\Psi\rangle = \mathcal{P}_G \mathcal{P}_J |\Phi_{\rm mf}\rangle,
    \label{eq:vmc_ansatz}
\end{equation}
where $\mathcal{P}_{G} = \exp\left[-\sum_{i\ell} g_{\ell}n_{i\ell\uparrow}n_{i\ell\downarrow}\right]$ is the Gutzwiller projector suppressing local double occupancy, and $\mathcal{P}_J = \exp\left[-\sum_{\ell}\sum_{m=1}^{3}v_{m\ell}\sum_{\langle ij\rangle_m}n_{i\ell}n_{j\ell}\right]$ is the intralayer Jastrow factor incorporating density-density correlations up to third-nearest neighbors. The parameters $g_{\ell}$ and $v_{m\ell}$ are optimized independently for inequivalent layers. The mean-field ansatz $|\Phi_{\rm mf}\rangle$ is the ground state of an auxiliary Hamiltonian $H_{\rm mf}$ (see Supplemental Material~\ref{appendix:vmc}), parameterized by adjacent-layer hoppings $\widetilde{\chi}_{\ell\ell^\prime}$, layer-dependent chemical potentials $\widetilde{\mu}_\ell$, and intralayer pairing fields $\eta_\ell^{s,d}$. For a comprehensive comparison, we examine three candidate mean-field ansatzes: a spin-singlet $d_{x^2-y^2}$-wave state, an extended $s$-wave state, and an unpaired reference state with all pairing fields set to zero.

The mean-field variational parameters, together with the Gutzwiller as well as Jastrow factors, $g_\ell$ and $v_{m\ell}$, are determined by minimizing the variational energy $\langle\Psi|\mathcal{H}|\Psi\rangle/\langle\Psi|\Psi\rangle$ using the stochastic reconfiguration method~\cite{vmc_sorella_1, vmc_sorella_2}. 
Numerical simulations are performed on a $20 \times 20 \times 5$ lattice, focusing on $22\%$ hole doping and $U/|t_{1o}|=8$. To ensure convergence to the true global minimum, the variational ground state is selected from optimizations initialized with several distinct mean-field configurations. Notably, we find that the $d$-wave superconducting state is energetically favored over the non-superconducting ansatz. The extended $s$-wave pairing is strongly suppressed, i.e., being roughly two orders of magnitude weaker than the $d$-wave amplitudes. 
Furthermore, the layer-resolved pairing amplitudes are found to be comparable in magnitude, and the electron density remains nearly uniform across all layers (with interlayer variations below $1\%$). 


Having established the \(d_{x^2-y^2}\)-wave ground state, we next examine its experimentally accessible signatures. The availability of bulk 6512 makes several probes of the superconducting order parameter feasible that are challenging for superconducting nickelate thin films. We therefore calculate the quasiparticle TDOS, superfluid density, spin Knight shift, and nuclear spin-lattice relaxation rate, which together probe the nodal structure and spin character of the superconducting state.

To evaluate these responses, we construct an effective Hamiltonian $\cH_{\rm eff}$ based on the VMC-optimized variational state. The effect of the Gutzwiller projection is incorporated approximately through the Gutzwiller approximation, equivalently formulated in terms of slave-boson renormalization factors~\cite{KRSB_01} determined from the local occupation probabilities measured in VMC. The main effect is to renormalize the intra- and interlayer hopping amplitudes according to
$t_{m\ell}\rightarrow z_\ell^2 t_{m\ell},$ and $\widetilde{\chi}_{\ell\ell^\prime} \rightarrow z_\ell z_{\ell^\prime}\widetilde{\chi}_{\ell\ell^\prime}$,
where the layer-dependent factor $z_\ell$ is
\begin{equation}
    z_\ell =
    \frac{
    \sqrt{n_{\ell h}n_{\ell\uparrow}}
    +
    \sqrt{n_{\ell\downarrow}n_{\ell d}}
    }
    {\sqrt{(1-n_\ell/2)(n_\ell/2)}}.
\end{equation}
Here $n_\ell$ is the electron density per site on layer $\ell$ and $n_{\ell\uparrow}=n_{\ell\downarrow}$ denotes the probability of single occupation with spin up or down. $n_{\ell d}$ and $n_{\ell h}$ denote the probabilities of double occupation and an empty site, respectively. The layer-dependent chemical potentials in $\mathcal H_{\rm eff}$ are adjusted to reproduce the electron densities. 

\begin{figure}
    \centering
    \includegraphics[width=\linewidth]{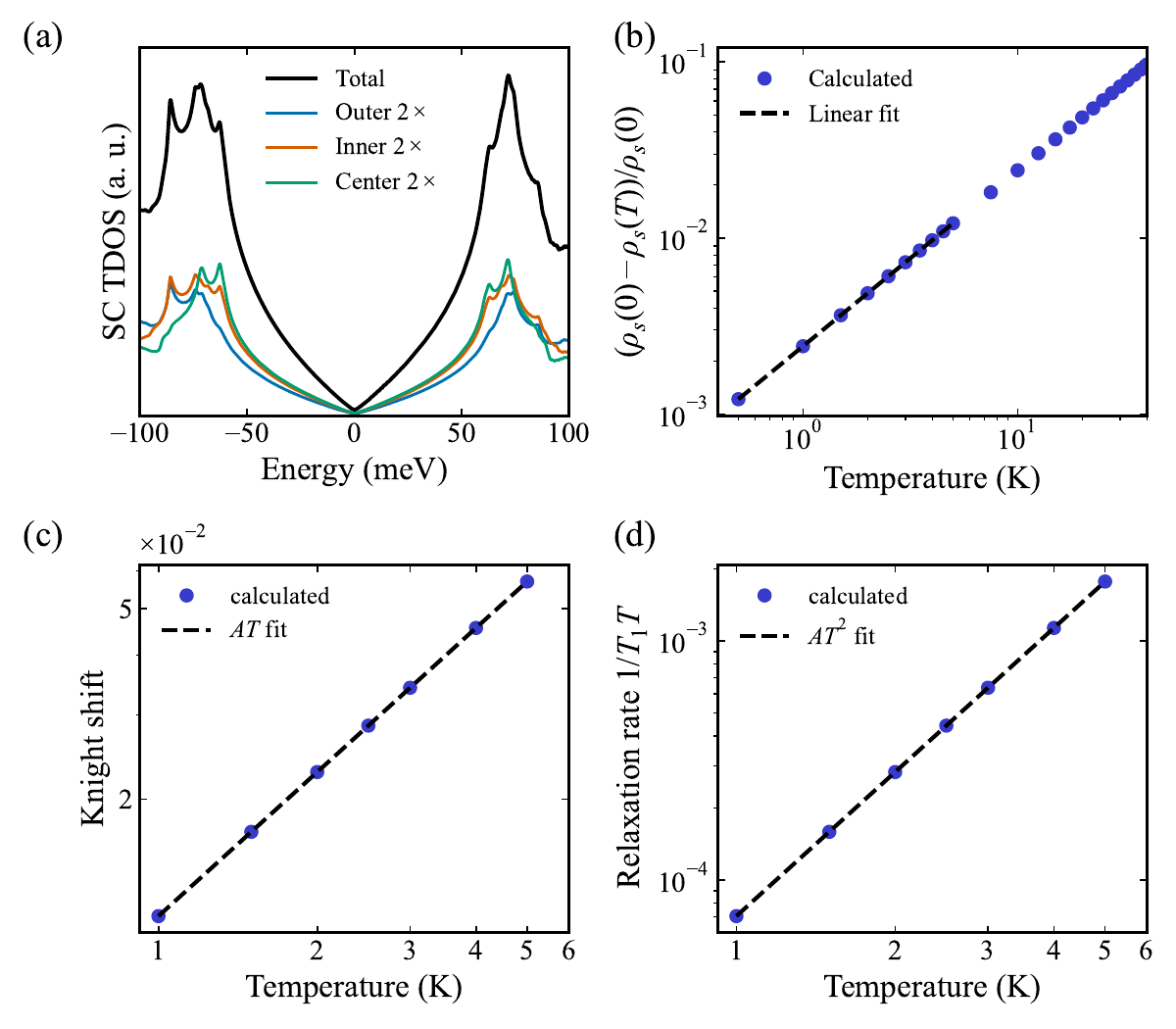}
    \caption{(a) Tunneling density of states of the superconducting state obtained from the renormalized effective Hamiltonian $\cH_{\rm eff}$. The individual-layer contributions are multiplied by two for visual clarity. (b) Temperature dependence of the normalized in-plane superfluid-density depletion $[\rho_s(0)-\rho_s(T)]/\rho_s(0)$, where $\rho_s=(\rho_{s,x}+\rho_{s,y})/2$. Symbols represent the calculated results, and the dashed line shows a linear fit over $0.5$--$5$ K. (c) Knight shift $K_s$ and (d) spin-lattice relaxation rate $1/T_1T$. Symbols represent the calculated results, and dashed lines show fits to $K_s\propto T$ and $1/T_1T\propto T^2$, respectively, over $1$--$5$ K.}
    \label{fig:property}
\end{figure}

We first consider the quasiparticle TDOS. Fig.~\ref{fig:property}(a) shows the layer-summed TDOS, $N_{\rm tot}(E)=2N_1(E)+2N_2(E)+N_3(E)$,
together with the outer-, inner-, and center-layer contributions \(N_1(E)\), \(N_2(E)\), and \(N_3(E)\), respectively. The momentum integration is performed on a \(2000\times2000\) mesh with a Lorentzian broadening \(\Gamma=0.8\) meV. At sufficiently low temperature, the TDOS is directly related to the differential tunneling conductance measured by scanning tunneling spectroscopy. The calculated spectrum exhibits a pronounced V-shaped suppression at low energy, reflecting the linearly vanishing density of states associated with the \(d\)-wave nodes. Additional features near the coherence peaks originate from van Hove singularities inherited from the normal-state band structure. The layer-summed TDOS therefore provides a direct reference for tunneling experiments without assuming layer-dependent tunneling matrix elements.

A second direct probe of the nodal quasiparticles is the superfluid density. We calculate the in-plane superfluid stiffness defined as $\rho_s(T)=[\rho_{s,x}(T)+\rho_{s,y}(T)]/2$ from the static transverse electromagnetic response, with details given in Supplemental Material~\ref{appendix:sd}. As shown in Fig.~\ref{fig:property}(b), the normalized low-temperature depletion $[\rho_s(0)-\rho_s(T)]/\rho_s(0)$ is linear in \(T\). This behavior arises from thermally excited quasiparticles near the gap nodes and contrasts with the exponentially activated response expected for a fully gapped superconductor. Penetration-depth measurements, which directly access this temperature dependence, therefore provide a particularly sensitive bulk test of the predicted nodal state. Similar measurements in infinite-layer nickelate films have already placed important constraints on the superconducting gap structure~\cite{2025Hwangi,2023Ariando}, and have also been proposed as a probe of pairing symmetry in La$_3$Ni$_2$O$_7$~\cite{2025Zhangh}.

NMR provides complementary information on the spin structure of the Cooper pairs \cite{NMR_d}. We calculate both the spin Knight shift and the nuclear spin-lattice relaxation rate using the same renormalized Hamiltonian, assuming a momentum-independent hyperfine form factor and equal coupling to all five layers. The spin contribution to the Knight shift is proportional to the uniform static susceptibility, $K_s(T)\propto A_\parallel(\textbf{0})\chi_\parallel(\textbf{0},0;T)$. For a field applied along \(z\), we define $\chi_\parallel=\chi_{zz}$ and $\chi_\perp=(\chi_{xx}+\chi_{yy})/2$, which are identical in the spin-rotation-invariant model considered here. As shown in Fig.~\ref{fig:property}(c), \(K_s\) decreases below \(T_c\) and follows an approximately linear temperature dependence at low \(T\), consistent with spin-singlet pairing and the finite population of nodal quasiparticles. Finally, the nuclear spin-lattice relaxation rate \(1/T_1T\) probes low-energy spin fluctuations in the superconducting state.
At low temperatures, relaxation is dominated by quasiparticles near the \(d\)-wave nodes. We obtain 
$1/T_1T\propto T^2$, as depicted in Fig.~\ref{fig:property}(d),
equivalent to the familiar \(1/T_1\propto T^3\) behavior of a clean nodal superconductor.

We further investigate whether the pairing tendency found in 6512 extends to other reduced RP nickelates, focusing on the trilayer 438 and infinite-layer 112 compounds. Superconducting signatures have recently been reported in substituted 438 single crystals~\cite{Ahn2026}, while superconductivity is well established in hole-doped infinite-layer 112 thin films~\cite{Li2019,2026Lib}, with transition temperatures reaching \(40\) K reported recently~\cite{Chow2025, 2026Lib}. These developments motivate a comparative study of the pairing tendencies across different members of the layered nickelate family.

For the trilayer 438 compound, we employ a three-layer Hubbard model analogous to the five-layer model in Eq.~\eqref{eq:Hubbard}, with \(U/|t_{1o}|=8\). The TB parameters obtained from DFT are summarized in Supplemental Material~\ref{appendix:trilayer}. VMC calculations are performed on \(L\times L\times3\) lattices at the nominal hole concentration \(x=1/3\). 
We find that the system lies close to the boundary between a \(d\)-wave superconducting state and a non-superconducting state, whereas the optimized extended \(s\)-wave pairing amplitude is vanishingly small. This near-degeneracy suggests that reducing the hole concentration by electron doping may provide a promising route to stabilize \(d\)-wave superconductivity in 438.

For the infinite-layer 112 compound, we consider an effective single-layer Hubbard model~\cite{Nomura2019,Ryee2020} with \(U/|t_1|=8\) at a hole concentration of $20\%$. The TB parameters are taken from the in-plane hopping amplitudes of Ref.~\cite{Chen2025}, and the VMC calculations are performed on a $20\times20$ lattice.
The VMC calculations find the lowest-energy \(d\)-wave state is energetically favored over the non-superconducting state, while the extended \(s\)-wave solution exhibits a negligible pairing amplitude. These results support a robust \(d_{x^2-y^2}\)-wave pairing tendency in 112, consistent with that obtained for the quintuple-layer 6512 compound.

In summary, we have investigated superconductivity in the quintuple-layer 6512 nickelate using a realistic multilayer Hubbard model and variational Monte Carlo calculations. We find a spin-singlet \(d_{x^2-y^2}\)-wave superconducting state, while the competing extended \(s\)-wave channel is strongly suppressed. Based on the VMC-optimized state, we further calculate several experimentally accessible quantities and find a V-shaped TDOS, a linear low-temperature depletion of the superfluid density, a suppressed spin Knight shift, and a nuclear spin-lattice relaxation rate \(1/T_1T\propto T^2\), providing a consistent set of signatures of a clean nodal superconductor. The availability of bulk 6512 makes these predictions particularly relevant, as penetration-depth and NMR measurements can directly test the pairing state identified here. Extending the same framework to related reduced RP nickelates, we find a clear \(d\)-wave tendency in infinite-layer 112, while trilayer 438 at its nominal \(1/3\) hole concentration lies close to the boundary between \(d\)-wave superconductivity and a non-superconducting state, suggesting electron doping as a possible route to enhance superconductivity.

Taken together, these results point to a common \(d_{x^2-y^2}\)-wave pairing tendency across the reduced RP nickelate family, with the strength of superconductivity controlled by carrier concentration and layer structure. Bulk 6512 therefore provides not only a new platform for establishing the pairing symmetry of nickelate superconductors, but also a bridge between superconducting infinite-layer 112 and the emerging superconducting regime of finite-layer compounds such as 438.

{\em Acknowledgments.}---We acknowledge stimulating discussions with Prof. Xianhui Chen, Tao Wu, and Zhenyu Wang. We acknowledge the support by the National Natural Science Foundation of China (Grant NSFC-12494594, NSFC-12574150, NSFC-12174428, NSFC-12504180), the Ministry of Science and Technology (Grant No. 2022YFA1403900), the Chinese Academy of Sciences Project for Young Scientists in Basic Research (2022YSBR-048), the New Cornerstone Investigator Program, and the start-up funding from ShanghaiTech University.

\bibliographystyle{apsrev4-2}
\bibliography{5-layer-Ni}

\clearpage

\appendix
\onecolumngrid

\begin{center}
\textbf{\large Supplemental Material: $d$-wave superconductivity in quintuple-layer bulk superconductor \LNO}
\end{center}

\appendix
\renewcommand{\appendixname}{Supplemental Material}
\setcounter{secnumdepth}{2}
\renewcommand{\thesection}{S\arabic{section}}

\pagenumbering{arabic}
\renewcommand{\thepage}{S\arabic{page}}

\setcounter{figure}{0}
\renewcommand{\thefigure}{S\arabic{figure}}

\setcounter{equation}{0}
\renewcommand{\theequation}{S\arabic{equation}}

\setcounter{table}{0}
\renewcommand{\thetable}{S\arabic{table}}

\section{DFT Computational Method}\label{appendix:dft}
Our first-principles calculations were performed within the framework of density functional theory (DFT), as implemented in the Vienna \textit{ab initio} Simulation Package (VASP) \cite{Kresse1996,Kresse1996PRB}. The projector augmented-wave (PAW) method was employed to describe the electron-ion interactions \cite{Blochl1994, Kresse1999}. The exchange--correlation energy was treated using the generalized gradient approximation (GGA) in the Perdew--Burke--Ernzerhof (PBE) parametrization \cite{PBE1996}. The plane-wave kinetic-energy cutoff was set to 500~eV, and the Brillouin zone was sampled using a 7 $\times$ 7 $\times$ 7 $k$-point mesh. The convergence criterion for the electronic self-consistent calculations was set to $10^{-8}$~eV.

\section{Variational wave function and optimization}
\label{appendix:vmc}

The mean-field ansatz $|\Phi_{\rm mf}\rangle$ entering Eq.~\eqref{eq:vmc_ansatz} is the ground state of the free-fermion Hamiltonian:
\begin{align}
    H_{\rm mf} =& \sum_{\bk,\sigma}\sum_{\ell=1}^{5}\left[\varepsilon^0_{\ell}(\bk)-\widetilde{\mu}_{\ell}\right]c_{\bk\ell\sigma}^\dagger c_{\bk\ell\sigma}\notag\\
    &+\sum_{\bk,\sigma}\sum_{\ell=1}^{4}\widetilde{\chi}_{\ell,\ell+1}(\bk)\left(c_{\bk\ell\sigma}^\dagger c_{\bk,\ell+1,\sigma}+\text{h.c.}\right)\notag\\
    &+\sum_{\bk}\sum_{\ell=1}^{5}\left[\widetilde{\Delta}_{\ell}(\bk)c_{\bk\ell\uparrow}^\dagger c_{-\bk\ell\downarrow}^\dagger+\text{h.c.}\right].
    \label{eq:Hmf}
\end{align}
Here $\varepsilon^0_\ell(\bk)=\sum_{m=1}^3t_{m\ell}\gamma_{m\bk}$ denotes the physical intralayer dispersion. The auxiliary layer-dependent chemical potentials $\widetilde{\mu}_\ell$ are optimized independently for the three layer types. The auxiliary interlayer matrix elements retain the physical momentum dependence,
\begin{equation}
    \widetilde{\chi}_{\ell,\ell+1}(\bk)=-2\widetilde{\chi}_{\ell,\ell+1}(\cos k_x-\cos k_y)^2.
\end{equation}
Mirror symmetry imposes $\widetilde{\chi}_{12}=\widetilde{\chi}_{45}=\widetilde{\chi}_{oi}$ and $\widetilde{\chi}_{23}=\widetilde{\chi}_{34}=\widetilde{\chi}_{ic}$, together with identical intralayer variational parameters on layers $1,5$ and on layers $2,4$.

For the two paired ansatzes, the nearest-neighbor spin-singlet pairing fields are
\begin{align}
    \widetilde{\Delta}^{d}_\ell(\bk)&=2\eta^{d}_\ell(\cos k_x-\cos k_y),\\
    \widetilde{\Delta}^{s}_\ell(\bk)&=2\eta^{s}_\ell(\cos k_x+\cos k_y).
\end{align}
All variational parameters are real, and no explicit interlayer pairing field is included. The Gutzwiller projector multiplies each configuration amplitude by $\exp[-\sum_\ell g_\ell D_\ell]$, where $D_\ell=\sum_i n_{i\ell\uparrow}n_{i\ell\downarrow}$ counts the doubly occupied sites on layer $\ell$. The Jastrow factor $\mathcal{P}_J=\exp\left[-\sum_{\ell}\sum_{m=1}^{3}v_{m\ell}\sum_{\langle ij\rangle_m}n_{i\ell}n_{j\ell}\right]$ includes intralayer density correlations up to 3rd nearest neighbors, where $n_{i\ell}=n_{i\ell\uparrow}+n_{i\ell\downarrow}$.

Each paired ansatz for the five-layer model therefore contains twenty variational parameters: three auxiliary chemical potentials $\widetilde{\mu}_{\ell}$, three nearest-neighbor pairing amplitudes $\eta^{d,s}_\ell$, two interlayer hopping amplitudes $\widetilde{\chi}_{\ell,\ell+1}$, three Gutzwiller parameters $g_\ell$, and nine Jastrow coefficients $v_{m\ell}$. The normal reference contains seventeen variational parameters, with the pairing amplitudes fixed to zero.

The same variational strategy is applied to the trilayer and single-layer models. For the trilayer model, mirror symmetry leads to two inequivalent layer types and six Jastrow coefficients. Both outer-inner and outer-outer auxiliary hopping amplitudes are optimized, giving fourteen variational parameters for each paired ansatz and twelve for the non-superconducting reference. For the single-layer model, the optimized parameter set consists of the auxiliary chemical potential, the nearest-neighbor pairing amplitude, the Gutzwiller parameter, and three Jastrow coefficients. This gives six parameters for each paired ansatz and five for the non-superconducting reference.

We employ open boundary conditions in the out-of-plane direction. To mitigate finite-size effects, periodic boundary conditions are used in the in-plane directions. Note that to lift the accidental degeneracy of the energy levels, the hopping matrix elements of mean-field Hamiltonian crossing the $x$ and $y$ boundaries are scaled by factors of $0.83$ and $1.07$, respectively. Nevertheless, the pairing and Jastrow bonds remain unmodified across either boundary. Note that the same boundary condition is applied to the physical Hamiltonians for five-layer, tri-layer, and single-layer models.

\section{Calculation of the superfluid density}
\label{appendix:sd}

We calculate the in-plane superfluid density from the static transverse electromagnetic response of the renormalized effective mean-field Hamiltonian. In the superconducting state, the effective Hamiltonian takes the Bogoliubov–de Gennes (BdG) form. In the five-layer Nambu basis $\Psi_{\bk}=(c_{\bk 1\uparrow},\ldots,c_{\bk 5\uparrow},c^\dagger_{-\bk 1\downarrow},\ldots,c^\dagger_{-\bk 5\downarrow})^T$, the Hamiltonian matrix is
\begin{equation}
    \mathcal{H}_{\rm eff}(\bk)=
    \begin{pmatrix}
        h(\bk) & \Delta(\bk)\\
        \Delta^\dagger(\bk) & -h^T(-\bk)
    \end{pmatrix},
\end{equation}
where $h(\bk)$ contains the Kotliar--Ruckenstein (KR)-renormalized intralayer and interlayer hoppings, onsite energies, and density-matched layer-dependent chemical potentials. The pairing matrix $\Delta(\bk)$ contains the optimized $d$-wave pairing amplitudes. All parameters are held fixed as temperature varies. We denote the ten eigenvalues and normalized eigenvectors by $E_{n\bk}$ and $|n,\bk\rangle$, respectively, and write $|n,\bk\rangle=(u_{n\bk},v_{n\bk})^T$, where $u_{n\bk}$ and $v_{n\bk}$ are five-component electron and hole vectors.

The current response to an external vector potential is written as
\begin{equation}
    j_\mu(\bq,\omega)=-\sum_\nu K_{\mu\nu}(\bq,\omega)A_\nu(\bq,\omega),
\end{equation}
where $\mu,\nu=x,y$ denote the in-plane directions and $K_{\mu\nu}$ is the electromagnetic response kernel. The superfluid stiffness is obtained from its static transverse limit,
\begin{equation}
    \rho_{s,\mu}(T)=\lim_{\substack{\bq\rightarrow0\\ \bq\perp\hat{\boldsymbol e}_\mu}}\operatorname{Re}K_{\mu\mu}(\bq,\omega=0;T).
    \label{eq:sd_transverse}
\end{equation}
We drop the common electromagnetic and geometrical factors, which cancel in the normalized stiffness reported in the main text. Momentum sums below are normalized per five-layer in-plane unit cell.

The response kernel consists of diamagnetic and paramagnetic contributions,
\begin{equation}
    K_{\mu\nu}(\bq,\omega)=D_{\mu\nu}+\Pi_{\mu\nu}(\bq,\omega).
\end{equation}
Introducing the vector potential through Peierls phases in the hopping terms gives the kinetic velocity and curvature matrices
\begin{equation}
    v_\mu(\bk)=\frac{\partial h(\bk)}{\partial k_\mu},
    \qquad
    w_{\mu\nu}(\bk)=\frac{\partial^2 h(\bk)}{\partial k_\mu\partial k_\nu}.
    \label{eq:sd_vertices}
\end{equation}
These derivatives include the momentum dependence of both intralayer and interlayer hoppings. The pairing matrix is held fixed in constructing the electromagnetic vertices. For the real, inversion-symmetric normal-state matrix used here, $h(\bk)=h^T(\bk)=h(-\bk)$, the current vertex in Nambu space is
\begin{equation}
    \mathcal{J}_\mu(\bk)=
    \begin{pmatrix}
        v_\mu(\bk) & 0\\
        0 & v_\mu(\bk)
    \end{pmatrix}.
    \label{eq:sd_current_vertex}
\end{equation}

The diamagnetic contribution is determined by the superconducting-state electronic density matrix,
\begin{equation}
    D_{\mu\nu}(T)=\frac{1}{N_{\bk}}\sum_{\bk,\sigma}\sum_{\ell,\ell^\prime=1}^{5}[w_{\mu\nu}(\bk)]_{\ell\ell^\prime}\langle c^\dagger_{\bk\ell\sigma}c_{\bk\ell^\prime\sigma}\rangle.
\end{equation}
Using the BdG eigenvectors and spin symmetry, this becomes
\begin{equation}
    D_{\mu\nu}(T)=\frac{2}{N_{\bk}}\sum_{\bk}\sum_{n=1}^{10}f(E_{n\bk})\,u^\dagger_{n\bk}w_{\mu\nu}(\bk)u_{n\bk},
    \label{eq:sd_diamagnetic}
\end{equation}
where $f(E)=[\exp(\beta E)+1]^{-1}$ and $\beta=(k_{\rm B}T)^{-1}$. The factor of two accounts for spin degeneracy.

The paramagnetic contribution is evaluated from the current-current correlation function. The Nambu Green's function has the spectral representation
\begin{equation}
    \mathcal{G}(\bk,i\omega_m)=\left[i\omega_m I_{10}-\mathcal{H}_{\rm eff}(\bk)\right]^{-1}=\sum_{n=1}^{10}\frac{|n,\bk\rangle\langle n,\bk|}{i\omega_m-E_{n\bk}},
\end{equation}
where $\omega_m=(2m+1)\pi/\beta$ is a fermionic Matsubara frequency. In the long-wavelength limit, the current-current response is
\begin{equation}
    \Pi_{\mu\nu}(\bq,i\Omega_\ell)=\frac{1}{\beta N_{\bk}}\sum_{\bk,m}\operatorname{Tr}\left[\mathcal{G}(\bk+\bq,i\omega_m+i\Omega_\ell)\mathcal{J}_\mu(\bk+\bq/2)\mathcal{G}(\bk,i\omega_m)\mathcal{J}_\nu(\bk+\bq/2)\right],
    \label{eq:sd_current_bubble}
\end{equation}
with $\Omega_\ell=2\ell\pi/\beta$ a bosonic Matsubara frequency and the trace taken over the ten Nambu components. We set the external frequency to zero before taking the transverse long-wavelength limit.

Substituting the spectral representation into Eq.~\eqref{eq:sd_current_bubble}, the Matsubara sum is evaluated using
\begin{equation}
    \frac{1}{\beta}\sum_m\frac{1}{(i\omega_m-E_a)(i\omega_m-E_b)}
    =\mathcal{F}_T(E_a,E_b),
\end{equation}
where
\begin{equation}
    \mathcal{F}_T(E_a,E_b)=
    \begin{cases}
        \displaystyle\frac{f(E_a)-f(E_b)}{E_a-E_b}, & E_a\neq E_b,\\[6pt]
        f'(E_a), & E_a=E_b.
    \end{cases}
    \label{eq:sd_divided_difference}
\end{equation}
The equal-energy limit is retained for both diagonal terms and degenerate states. The resulting static transverse paramagnetic contribution is
\begin{equation}
    \Pi_{\mu\mu}^{\rm tr}(T)=\frac{1}{N_{\bk}}\sum_{\bk}\sum_{n,m=1}^{10}\mathcal{F}_T(E_{n\bk},E_{m\bk})\left|\langle n,\bk|\mathcal{J}_\mu(\bk)|m,\bk\rangle\right|^2.
    \label{eq:sd_paramagnetic}
\end{equation}
Both positive- and negative-energy BdG states are included, and all intraband and interband current matrix elements are retained.

Combining Eqs.~\eqref{eq:sd_diamagnetic} and \eqref{eq:sd_paramagnetic}, the superfluid stiffness is
\begin{equation}
    \rho_{s,\mu}(T)=D_{\mu\mu}(T)+\Pi_{\mu\mu}^{\rm tr}(T).
\end{equation}
We report the in-plane average $\rho_s(T)=[\rho_{s,x}(T)+\rho_{s,y}(T)]/2$ and its normalized depletion $[\rho_s(0)-\rho_s(T)]/\rho_s(0)$.

The Brillouin-zone integrals are evaluated numerically, with locally refined momentum meshes around all superconducting nodes. For calculations down to $0.5$ K, we use a $2000\times2000$ background grid, a local sampling density equivalent to $24001\times24001$ within radius $0.20$ of the nodes, and a further refinement equivalent to $96001\times96001$ within radius $0.03$.

\section{Tight-binding model and parameters for 438}
\label{appendix:trilayer}

We describe the Ni $d_{x^2-y^2}$ sector of the trilayer compound La$_4$Ni$_3$O$_8$ by a two-dimensional three-band tight-binding (TB) model fitted to the DFT band dispersions. Following the notation used for 6512, the layers are ordered as outer(1)--inner(2)--outer(3), with $\ell=1,2,3$, respectively, and the two outer layers share the same in-plane hopping parameters. In the basis $C_{\bk\sigma}=(c_{\bk1\sigma},c_{\bk2\sigma},c_{\bk3\sigma})^T$, the kinetic Hamiltonian is
\begin{equation}
    \cH_t^{(438)}=\sum_{\bk,\sigma}C_{\bk\sigma}^\dagger h_{438}(\bk)C_{\bk\sigma},
    \qquad
    h_{438}(\bk)=
    \begin{pmatrix}
        \varepsilon_{o,\bk} & \tau_{oi,\bk} & \tau_{oo,\bk}\\
        \tau_{oi,\bk} & \varepsilon_{i,\bk} & \tau_{oi,\bk}\\
        \tau_{oo,\bk} & \tau_{oi,\bk} & \varepsilon_{o,\bk}
    \end{pmatrix}.
    \label{eq:438_hopping_matrix}
\end{equation}
Here $o$ and $i$ denote the outer and inner layer types. Notably, we retain both outer--inner hopping and hopping between the two outer layers within a trilayer block.

The intralayer dispersion includes hopping up to the third nearest neighbor,
\begin{equation}
    \varepsilon_{\ell,\bk}=\epsilon_\ell+\sum_{m=1}^{3}t_{m\ell}\gamma_{m\bk},
    \qquad \ell=o,i.
    \label{eq:438_intralayer}
\end{equation}
$\epsilon_\ell$ denotes the layer-dependent onsite energy shift. The three corresponding harmonics are defined as in the main text.

\begin{figure}
    \centering
    \includegraphics[width=0.5\linewidth]{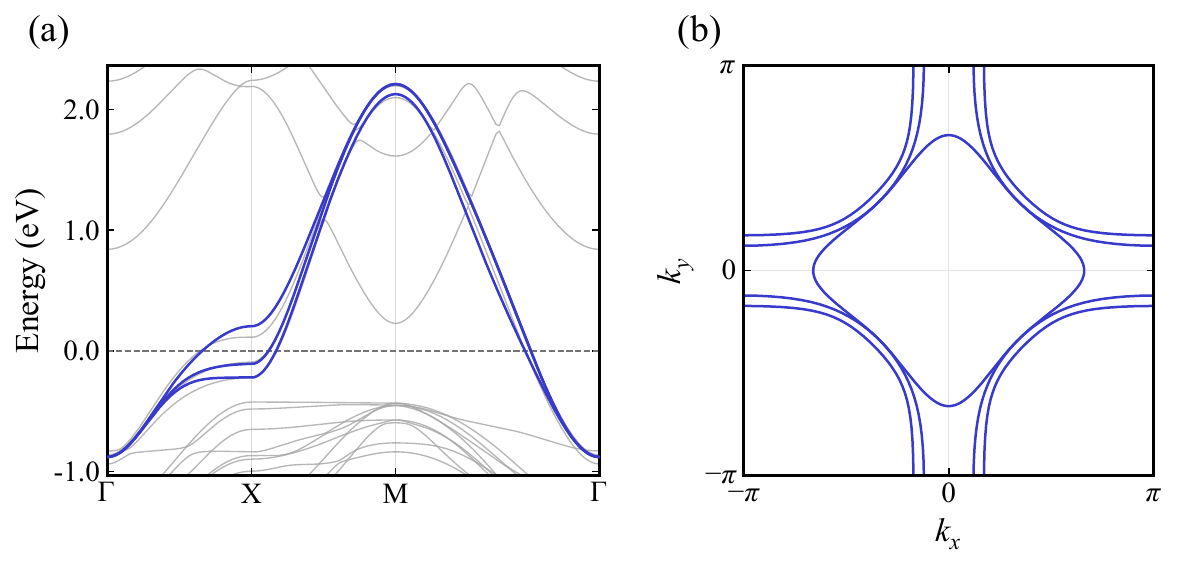}
    \caption{(a) Dispersion of the three-band TB model for 438 along $\Gamma$--$X$--$M$--$\Gamma$. The band dispersion obtained by DFT is depicted in light grey. (b) Corresponding Fermi surface. The total electron filling is two per trilayer unit cell, including spin degeneracy.}
    \label{fig:438_normal}
\end{figure}

Both interlayer matrix elements have the same form as in the main text,
\begin{equation}
    \tau_{\ell\ell^\prime,\bk}=-2t_\perp^{\ell\ell^\prime}(\cos k_x-\cos k_y)^2,
    \qquad \ell\ell^\prime=oi,oo.
    \label{eq:438_interlayer}
\end{equation} 
The $oo$ term couples layers 1 and 3 within the same block. The obtained dispersion and corresponding Fermi surface are depicted in Fig.~\ref{fig:438_normal}.

\begin{table}
    \centering
    \begin{tabular}{c|ccccc}
        \hline
        & $\epsilon_\ell$ & $t_{1\ell}$ & $t_{2\ell}$
        & $t_{3\ell}$ & $t_\perp^{\ell\ell^\prime}$ \\
        \hline
        outer
        & $0.631$ & $-0.369$ & $0.087$
        & $-0.048$ & $t_\perp^{oo}=-0.010$ \\
        inner
        & $0.542$ & $-0.353$ & $0.087$
        & $-0.028$ & $t_{\perp}^{oi}=0.018$ \\
        \hline
    \end{tabular}
    \caption{Tight-binding parameters of the trilayer 438
    model, in units of eV. The two entries in the last column denote outer--outer and outer--inner hopping, respectively.}
    \label{tab:TB438}
\end{table}

\end{document}